\documentclass{article}
\usepackage{spconf,amsmath,graphicx}
\usepackage{multirow}
\usepackage{amsmath,amsfonts}
\usepackage{bm}
\usepackage{subfigure}
 \usepackage{cite}

\title{DISENTANGLED FEATURE LEARNING FOR REAL-TIME NEURAL SPEECH CODING}
\name{Xue Jiang$^{1 \ast}$, Xiulian Peng$^{2}$, Yuan Zhang$^{1}$, Yan Lu$^{2}$\thanks{$^{\ast}$This work was done at Microsoft Research Asia.}}
\address{$^{1}$  Communication University of China, Beijing, China \\
$^{2}$  Microsoft Research Asia, Beijing, China}
\begin{document}
%
\maketitle
\begin{abstract}
Recently end-to-end neural audio/speech coding has shown its great potential to outperform traditional signal analysis based audio codecs. This is mostly achieved by following the VQ-VAE paradigm where blind features are learned, vector-quantized and coded. In this paper, instead of blind end-to-end learning, we propose to learn disentangled features for real-time neural speech coding. Specifically, more global-like speaker identity and local content features are learned with disentanglement to represent speech. Such a compact feature decomposition not only achieves better coding efficiency by exploiting bit allocation among different features but also provides the flexibility to do audio editing in embedding space, such as voice conversion in real-time communications. Both subjective and objective results demonstrate its coding efficiency and we find that the learned disentangled features show comparable performance on any-to-any voice conversion with modern self-supervised speech representation learning models with far less parameters and low latency, showing the potential of our neural coding framework.   
\end{abstract}
\section{Introduction}
\label{sec:intro}
Recently deep learning-based audio/speech codecs have made significant progress to deliver a high quality at very low bitrates \cite{wavcodec,sampleRNNcodec,Lyra,VQ-VAE-wavenet,soundstream,TFNetCodec,zhen2021scalable}. They are either based on strong generative models for decoding \cite{wavcodec,sampleRNNcodec,Lyra} or end-to-end learning by the VQ-VAE\cite{VQ-VAE} framework \cite{VQ-VAE-wavenet,soundstream,TFNetCodec}. However, the latent features to quantize are mostly blindly learned using a convolutional neural network (CNN) without any prior knowledge and therefore they are lack of interpretability. In this paper, we investigate how the disentangled feature learning can help for neural speech coding under the VQ-VAE paradigm. 

Several works have explored the disentanglement of speaker identity and content under the VQ-VAE paradigm. It is shown in \cite{VQ-VAE} that the vector quantization (VQ) space can learn phones and sub-phones when conditioned on a global speaker embedding at the decoder. \cite{disentangle} proposed a semi-supervised VQ-VAE to learn the disentangled phone and speaker representations. Such a speaker-content disentanglement is even more widely explored in voice conversion tasks \cite{VQVC,vqmivc,adain}, where VQ is usually performed as a strong information bottleneck to restrict information flow. VQVC\cite{VQVC} use VQ to disentangle content information and represent the speaker information by the difference between continuous latent and the discrete codes.
However, these works all focus on representation learning for downstream tasks like phoneme recognition, voice conversion, and so on. 
They pay no attention on speech coding and the bitrate. \cite{facebook} introduced disentangled representations for speech coding with large pretrained self-supervised learning (SSL) models but the pretrained content model is not explicitly disentangled with speaker identity and the speaker feature is globally learned without adaptation to general audio with multiple speakers.

\begin{figure}[tb]
\centering
\includegraphics[width=0.95\linewidth,height=0.45\linewidth]{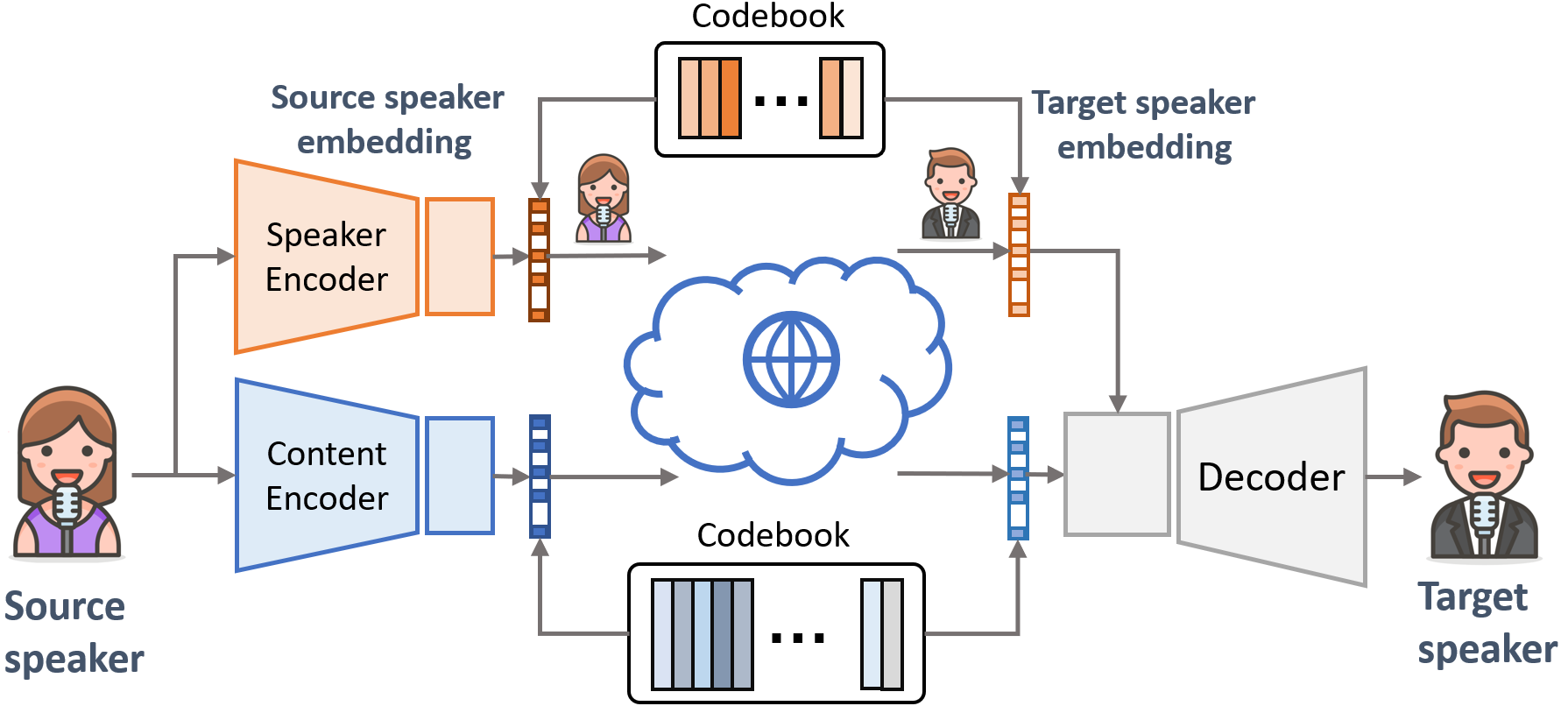}
\vspace{-0.5cm}
\caption{Disentangled speech coding and voice conversion for real-time communications.}
\vspace{-0.2cm}
\label{fig:system}
\vspace{-0.35cm}
\end{figure} 

\begin{figure*}[t]
\centering
\vspace{-1cm}
\begin{minipage}[b]{0.6\linewidth}
    \subfigure[Network structure]{\includegraphics[width=1\linewidth,height=0.4\linewidth]{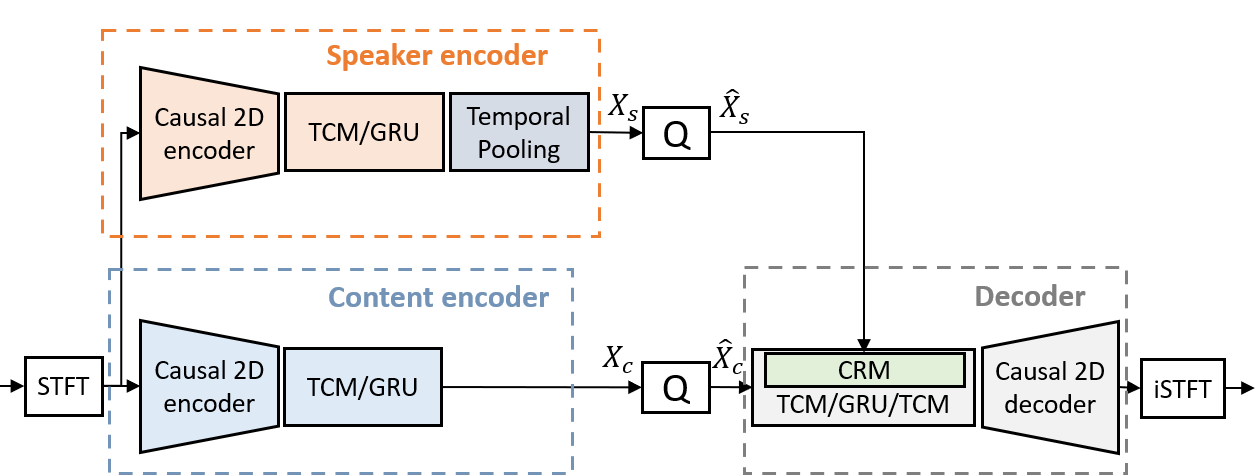}}
    \end{minipage}
\begin{minipage}[b]{0.33\linewidth}
    \subfigcapskip=-4.5pt
    \subfigbottomskip=0.1pt
    \subfigure[Conditional regularization module (CRM)]{\includegraphics[width=0.9\linewidth,,height=0.35\linewidth]{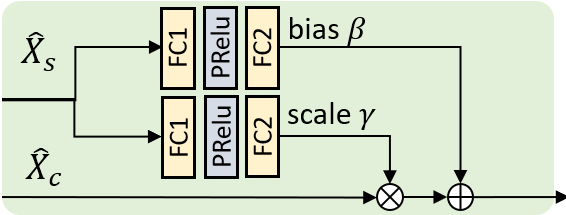}}
    \subfigure[Local causal pooling]{\includegraphics[width=0.9\linewidth,,height=0.4\linewidth]{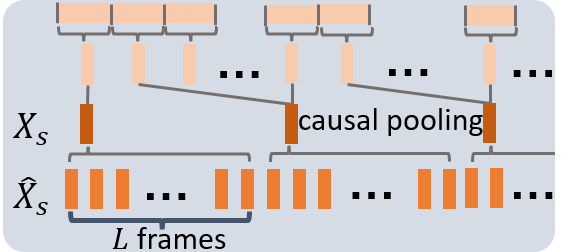}}
\end{minipage}
\vspace{-0.4cm}
\caption{The overall network structure of the proposed Disen-TF-Codec. }
\vspace{-0.4cm}
\label{fig:model structure}
\end{figure*}

In this paper, we discuss the disentanglement learning for neural speech coding under real-time communications. We propose the Disen-TF-Codec, a self-supervised low-latency neural speech codec that disentangles speaker with content. We investigate how the vector quantization and instance normalization help to remove speaker information from the content branch. We also study two encoding schemes for speaker branch, the \textit{global} for single speaker and the \textit{local} that can adapt to speaker change during communications. Automatic rate allocation between two branches is introduced to achieve good rate-distortion optimization. We show that such a disentanglement can help speech coding for extremely low-bitrate scenarios as more compact features are learned through disentanglement.

Moreover, the proposed framework naturally supports voice conversion during real-time communications by simply replacing the speaker embedding with the target one in compressed domain, as shown in Fig. \ref{fig:system}. When taking compression as a ``pretext'' task, we show that the learned content representations perform competitively with big SSL models on an any-to-any voice conversion benchmark with far less parameters and low latency.

\vspace{-0.3cm}
\section{The proposed scheme}
\vspace{-0.2cm}
\label{sec:proposed scheme}
\subsection{Overview}
\vspace{-0.2cm}
\label{ssec:overview}
As shown in Fig \ref{fig:model structure}(a), the proposed Disen-TF-Codec consists of two causal encoders to hold the speaker and content information flow, respectively. The content features are quantized through a learned vector quantizer and encoded with huffman coding. At decoding, the content features are fused with speaker information through a conditional regulation module (CRM) and several causal temporal filtering blocks. After a causal decoder, the speech is finally synthesized.

We consider two scenarios for real-time communications: (1) single speaker; (2) multiple speakers with speaker changes. In the first scenario, an enrollment step is needed to get a global speaker embedding during the first few seconds through global temporal pooling on speaker features. This global embedding can be transmitted just once so the speaker bitrate is quite low compared with content. In the second scenario, we leverage a local causal temporal pooling to get continuously updated speaker embeddings, which are quantized and transmitted every $L$ frames to adapt to speaker change. Automatic rate allocation between the two branches is introduced to balance bitrate with perceptual quality. All modules are trained from end to end and adversarial training is employed to get a good synthesized quality. In the following sections, we will describe these parts in detail.
\vspace{-0.3cm}
\subsection{Disentangled Feature Learning}
Existing studies \cite{VQ-VAE,vq-wav2vec} show that the discrete latents learned by VQ-VAE are highly correlated with phonemes, which indicates that VQ, as an information bottleneck, has the ability to disentangle speech information from content. Generally, a lower bitrate means a stronger information bottleneck, resulting in better disentanglement, but the audio quality is also sacrificed. In this paper, we investigate the disentanglement under two bitrates, i.e. 256bps and 1kbps and explore the coding efficiency and disentanglement ability of different training schemes. Details can be found in the experimental parts.

\vspace{-0.2cm}
\subsection{Network Structure}
\label{ssec:enc/dec}
We take the complex spectrum by short-time Fourier transform (STFT) as the network input, denoted by $X \in \mathbb{R}^{T\times F\times 2}$. 

\textbf{Content encoder}
Following the similar design as TFNet \cite{TFNetCodec}, the content encoder $E_c$ includes several 2D causal convolutional layers followed by causal temporal filtering modules. The 2D convolution layers capture local frequency and temporal correlations while the temporal filtering modules including TCM and GRU explore long-term dependencies from the past frames. The two-scale feature extraction helps to learn robust and powerful sequential representations of content information, denoted as $X_c \in \mathbb{R}^{T\times D_{c}}$. 

\textbf{Speaker encoder} The speaker encoder $E_{s}$ mostly takes similar structure as $E_{c}$ with a causal structure. For single speaker scenario, we introduce an additional global temporal average pooling layer to aggregate all frames into a single vector, followed by several fully-connected layers to generate a global speaker embedding $X_{s} \in \mathbb{R}^{1\times D_{s}}$. For the multi-speaker scenario, we introduce local causal temporal pooling to get continuously updated speaker embeddings $X_{s} \in \mathbb{R}^{T/L\times D_{s}}$ as shown in Fig. \ref{fig:model structure}(c). The features are aggregated every $L$ frames and quantized. During decoding, the embeddings are dequantized and each embedding is propagated to future $L$ frames to ensure causality and get the original temporal resolution.  In both schemes, the embeddings are extracted across a temporal range much larger than a phoneme duration thus no phoneme information is included in this branch.

\textbf{Decoder}
The decoder aims to reconstruct $\hat{X} \in \mathbb{R}^{T\times F\times 2}$ with the speaker and content information. It takes a mirror-like architecture of the content encoder $E_c$ but with more temporal filtering blocks for better reconstruction quality. A conditional regularization module (CRM) is introduced to merge the two branch information. Specifically, conditional information ($\bm{\gamma} $, $\bm{\beta}$) is learned from decoded speaker embeddings $\hat{X}_s$ and employed to regularize the decoded content representation $\hat{X}_c$, as given by
\vspace{-0.2cm}
\begin{equation}
\begin{aligned}
\bm{\gamma} = \mathcal{F}_1(\hat{X}_s),\bm{\beta} = \mathcal{F}_2(\hat{X}_s)\\
CRM(\hat{X}_c,\hat{X}_s) = \bm{\gamma} \times \hat{X}_c + \bm{\beta},
\end{aligned}
\vspace{-0.2cm}
\label{eq:merge module}
\end{equation}
where  $\bm{\gamma},\bm{\beta} \in \mathbb{R}^{1\times D_{c}} $ and $\bm{\gamma},\bm{\beta} \in \mathbb{R}^{T\times D_{c}}$ for \textit{global} and \textit{local} pooling cases represent the channel-wise scale and bias modulation parameters learned from $\hat{X}_s$ through linear projection layers $\mathcal{F}_{1,2}$.   We apply the same modulation parameter to all frames of $\hat{X}_c$ when \textit{global} speaker embedding is used and $\hat{X}_s$ is injected at multiple positions in the temporal filtering part of the decoding through CRM.


\vspace{-0.2cm}
\subsection{Vector Quantization and Rate Allocation}
\vspace{-0.2cm}
\label{ssec:vq}
Similar to our previous work \cite{predictivecodec}, we adopt the distance-gumbel-softmax-based scheme for vector quantization. During the forward pass, the codeword closer to the input will have a higher probability of being selected. During the backward pass, the gradient with respect to gumbel-softmax logits is used. 

Following \cite{predictivecodec}, entropy-based method is employed to constraint the total bitrate to a target value $R_{target}$. To estimate the entropy, we calculate the sample soft assignment distribution $\bm{Q}$ over $K$ codewords. The entropy of a quantized feature is then estimated by summing up the probabilistic assignment logits to each codeword within a minibatch, as given by
\vspace{-0.4cm}
\begin{equation}
\vspace{-0.15cm}
\begin{aligned}
    \mathcal{H}(\bm{Q}) &\approx -\sum_{k=1}^K Q_k\log Q_k.
\end{aligned}
\end{equation}
We allow automatic rate allocation between speaker and content features for the local pooling case. Therefore, the total bitrate is constraint by
\vspace{-0.2cm}
\begin{equation}
\vspace{-0.15cm}
\begin{aligned}
\mathcal{L}_{rate} &= ||R_{target} - \mathcal{H}(\bm{Q}_s) - \mathcal{H}(\bm{Q}_c)||_1,
\end{aligned}
\label{eq:L-rate}
\end{equation}
where $\mathcal{H}(\bm{Q}_s)$ and $\mathcal{H}(\bm{Q}_c)$ are entropy estimates for speaker and content features, respectively. Group vector quantization is employed to facilitate different bitrates. For codebook size, we do not encourage equal distribution of codewords as that in \cite{vq-wav2vec}. Instead we set a larger codebook and leverage the loss $\mathcal{L}_{rate}$ to control the total bitrate so that the real feature distribution can be captured.

\begin{table}[]
\vspace{-0.1cm}
\setlength\tabcolsep{2pt}
\renewcommand\arraystretch{0.9}
\centering
\caption{Evaluation of disentangled schemes at 256bps.  }\label{table:1}
\fontsize{7.5}{10}\selectfont
\begin{tabular}{|l|ccc|ccc|}
\hline
 \multirow{2}{*}{Method} & \multicolumn{3}{c|}{Reconstruction} & \multicolumn{3}{c|}{Voice Conversion} \\ \cline{2-7} 
                         & PESQ       & STOI      & VISQOL     & MCD        & WER        & ASV         \\ \hline
 TF-Codec                & 1.691      & 0.852     & 2.299      & -          & -          & -           \\ \cline{1-7} 
 Disen-TF-Codec (global)          & 1.867      & 0.873     & 2.520      & 8.60       & 22.6       & 64.5        \\ 
 Disen-TF-Codec (global w. IN)    & \textbf{1.954}      & \textbf{0.881}     &  \textbf{2.581}          & \textbf{8.57}       & \textbf{27.2 }      & \textbf{65.75 }      \\ \hline
\end{tabular}
\end{table}


\begin{table}[]
\vspace{-0.5cm}
\setlength\tabcolsep{2pt}
\renewcommand\arraystretch{0.9}
\centering
\caption{Evaluation of disentangled schemes at 1kbps (w/o adversarial training).  }\label{table:1}
\fontsize{7.5}{10}\selectfont
\begin{tabular}{|l|ccc|ccc|}
\hline
 \multirow{2}{*}{Method} & \multicolumn{3}{c|}{Reconstruction} & \multicolumn{3}{c|}{Voice Conversion} \\ \cline{2-7} 
                         & PESQ       & STOI      & VISQOL     & MCD        & WER        & ASV         \\ \hline
 TF-Codec    &   2.497    & 0.930     &  2.920  & -          & -  & -  \\ \cline{1-7}                                            
 Disen-TF-Codec (global) &   2.595     &   0.935       &  3.010          &  9.08          &     11.1         &    42.25         \\    
 Disen-TF-Codec (global, from 256)          &    2.478   &   0.927   &   2.905         &   8.90     &  9.1      &  50.00      \\ 
 Disen-TF-Codec (global w. IN)    &   \textbf{2.679}    &   \textbf{0.938}   &  \textbf{3.163}     &   \textbf{8.91}     &    \textbf{11.5}    &     \textbf{58.75}   \\  
 Disen-TF-Codec (local) &   2.547     &         0.931  &  2.945          &      9.24      &    13.2         &      37.25        \\ \hline
\end{tabular}
\vspace{-0.5cm}
\end{table}
\vspace{-0.1cm}
\section{Experimental Results}
\label{sec:exp}
\subsection{Dataset and Settings}
We take the train-clean-100 subset from LibriSpeech\cite{librispeech} as our training data, which covers 251 speakers. Each audio is cut into 3-second segments for training. For evaluation, we use Librispeech test-clean. Hanning window is used in STFT with a window length of 40 ms and a hop length of 10 ms. We optimize our network in an end-to-end fashion with multiple loss terms and adversarial training, similar to that in \cite{predictivecodec}.

We evaluate the proposed Disen-TF-Codec from two aspects, the coding efficiency and disentanglement ability. For reconstruction quality, we use PESQ, STOI and VISQOL as the evaluation metrics for ablation study and a subjective listening test for comparison with other codecs. For disentanglement, we leverage the voice conversion benchmark, S3PRL-VC \cite{superb-vc}, an extension of the SUPERB \cite{superb} toolkit that focuses on the voice conversion (VC) downstream task, to evaluate the discrete content features. S3PRL-VC assesses the converted audio by MCD, WER and ASV from the aspects of signal reconstruction, intelligibility and speaker similarity, respectively. The more challenging any-to-any (A2A) VC setting is chosen for comparison, with Taco2-AR as the downstream model. 


\vspace{-0.2cm}
\subsection{Ablation Study on Disentangled Schemes}
In this section, we compare the proposed Disen-TF-Codec with several variants to show its effectiveness. As shown in Table 1 and 2, the proposed schemes under two scenarios are denoted as ``Disen-TF-Codec (global)'' by global pooling and ``Disen-TF-Codec (local)'' by local causal pooling, respectively. We first compare with the single-branch baseline ``TF-Codec'' that uses no disentanglement at all. We can see that both two Disen-TF-Codec schemes outperform the ``TF-Codec'' in reconstruction quality, indicating that disentanglement helps for the coding efficiency for making better use of the limited bit budget. At 256bps, the gain is much larger in reconstruction quality with better ASV in voice conversion, showing that lower bitrate serves as a better information bottleneck to remove speaker information. Moreover, the ``Disen-TF-Codec (local)'' is slightly worse than ``Disen-TF-Codec (global)'' as more bits are allocated to the redundant speaker embeddings in ``Disen-TF-Codec (local)''. In our experiments, we found that the speaker branch takes 10\% of the total bitrate while 90\% bits are consumed by the content branch to deliver the key information in speech. 

We then compare with ``Disen-TF-Codec (global w. IN)'' that further introduces the instance normalization (IN) technique, that is widely used to remove speaker information in voice conversion \cite{adain}, for further disentanglement based on the ``Disen-TF-Codec (global)'' backbone. As IN is non-causal, we investigate it only on the single-speaker scenario where in enrollment step the mean/variance parameters of each IN layer at both encoder and decoder can be obtained similar to the global speaker embedding. The IN parameters of the decoder needs to be encoded into the bitstream with the global speaker embedding, the bitrate of which is negligible compared with content features. We can observe that Disen-TF-Codec with IN shows better reconstruction quality and voice conversion performance and the gap gets larger as the bitrate increases from 256bps to 1kbps. This indicates that as bitrate increases to 1kbps, VQ as the only information bottleneck is not strong enough and the discrete codes would still contains some speaker information; therefore IN plays as a key role to perform information constraint.

\begin{figure}
\centering
\vspace{-0.8cm}
\begin{minipage}[b]{0.6\linewidth}
    {\subfigure[Subjective listening test. ]{\includegraphics[width=1\linewidth,height=1.05\linewidth]{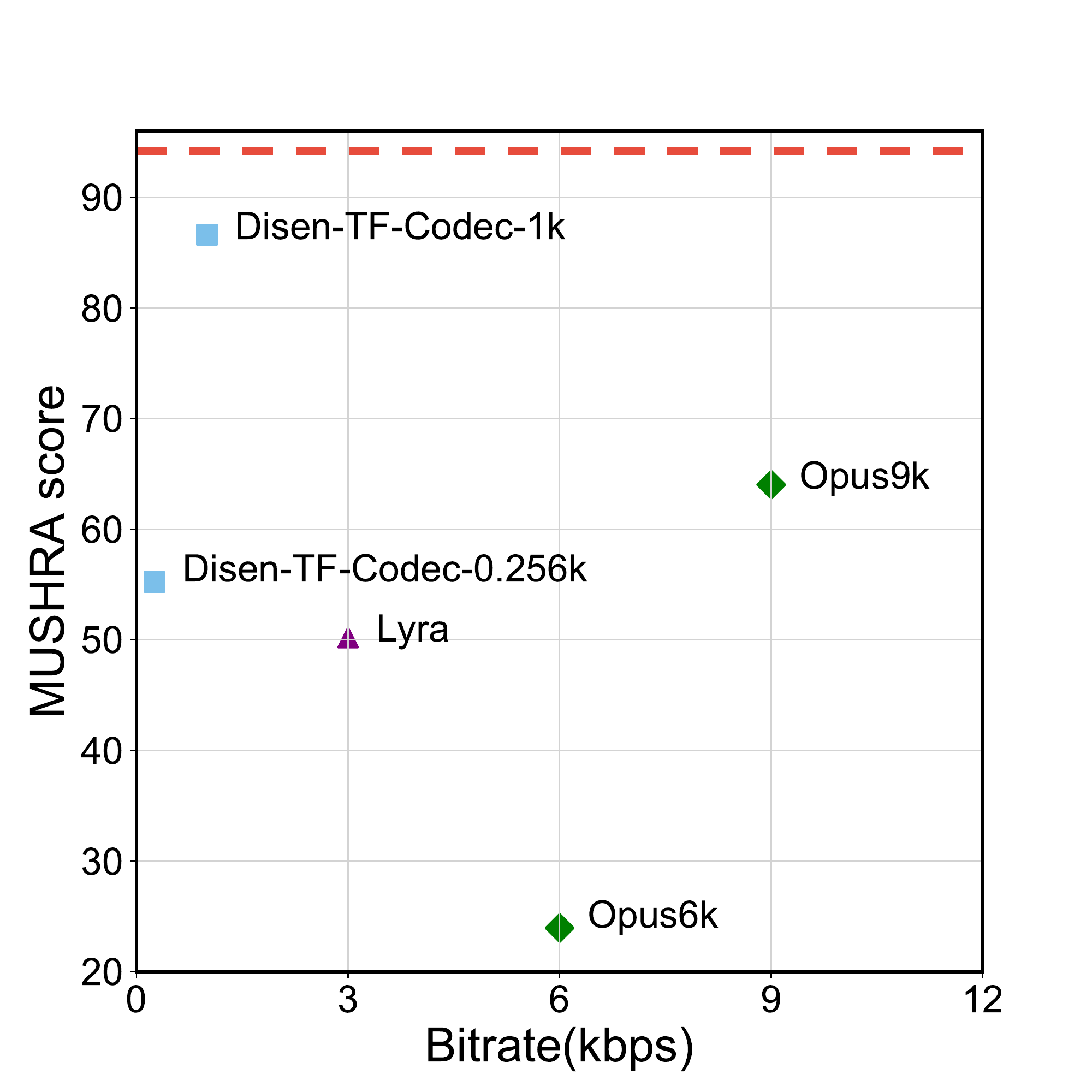}}}
\end{minipage}
\begin{minipage}[b]{0.28\linewidth}
    \subfigcapskip=-4.5pt
    \subfigbottomskip=0.1pt
    \subfigure[$X_s$]
    {\includegraphics[width=0.85\linewidth,height=0.65\linewidth]{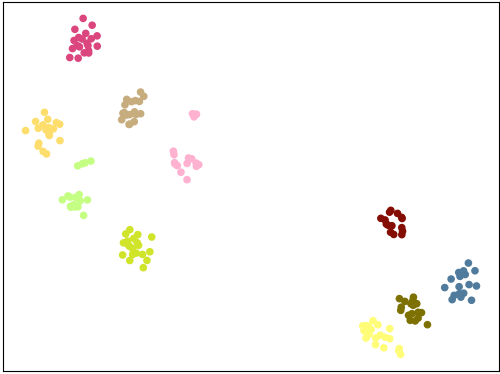}}
    \subfigure[ $X_c$]
    {\includegraphics[width=0.85\linewidth,height=0.65\linewidth]{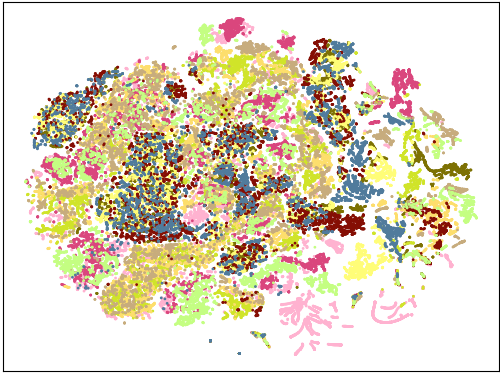}}
    \subfigure[ $\hat{X}_c$]
    {\includegraphics[width=0.85\linewidth,height=0.65\linewidth]{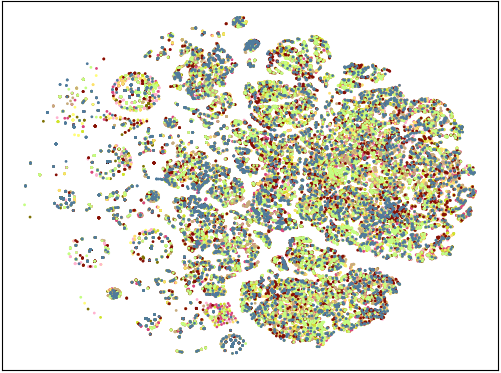}}
\end{minipage}
\vspace{-0.2cm}
\caption{(a) Disen-TF-Codec vs. standard codecs. The red dotted line represents the score of the reference. (b)(c)(d) t-SNE visualization of speaker representation$X_s$, content representations before and after VQ, i.e. $X_c$ and $\hat{X}_c$. Different colors represent different speakers.}
\vspace{-0.6cm}
\label{fig:mushra}
\end{figure}

For 1kbps, we also compare with a two-step training algorithm ``Disen-TF-Codec (global, from 256)'' that takes speaker and content features from 256bps for encoding at 1kbps where the content encoder is freezed without finetuning and the speaker encoder is finetuned. As shown in Table 2, this scheme preserves the high disentanglement brought by 256bps but sacrifices coding efficiency at 1kbps.

To look deep into the representations it learns, we perform t-SNE \cite{tsne} on $X_s$, $X_c$ and $\hat{X}_c$ for the ``Disen-TF-Codec (global)'' scheme at 256bps. 20 unseen speakers are used in this visualization. As shown in Fig \ref{fig:mushra}(b)(c)(d), the speaker features are clustered well for each speaker. The content features before VQ, $X_c$, still show some clustering, whereas for $\hat{X}_c$, the points scatter for most speakers. It indicates that without any explicitly speaker-related supervision, the speaker encoder learns good speaker-related information and the speaker information is effectively removed from $\hat{X}_c$ by VQ.


\vspace{-0.2cm}
\subsection{Comparison with Other Codecs}
\label{ssec:exp1}
We conduct a MUSHRA \cite{mushra} subjective listening test to measure the quality of the proposed codec, where 8 participants evaluate 12 samples. We compare our Disen-TF-Codec for single speaker case with Opus \cite{opus} and Lyra \cite{Lyra}, two codecs used for real-time communications. Fig.\ref{fig:mushra}(a) shows the subjective evaluation results. It is observed that Disen-TF-Codec at 256bps outperforms Lyra at 3kbps and Disen-TF-Codec at 1kbps achieves a high score which exceeds Opus 9kbps by a large margin. These results demonstrate its superiority. 
\begin{table}[]
\vspace{-1.1cm}
\caption{Results on any-to-any VC over various upstreams. }\label{table:2}
\centering
\renewcommand\arraystretch{0.85}
\fontsize{7.5}{10}\selectfont
\begin{tabular}{|l|l|lll|}
\hline
\multirow{2}{*}{Upstream} & \multirow{2}{*}{Params} & \multicolumn{3}{l|}{Intra-lingual A2A}                         \\ \cline{3-5} 
                          &                         & \multicolumn{1}{l|}{MCD}  & \multicolumn{1}{l|}{WER}  & ASV   \\ \hline
mel                       & -                       & \multicolumn{1}{l|}{9.49} & \multicolumn{1}{l|}{4.2}  & 19.50 \\ \hline
APC                       & 4.11M                   & \multicolumn{1}{l|}{9.57} & \multicolumn{1}{l|}{3.5}  & 23.25 \\ 
wav2vec                   & 32.54M                  & \multicolumn{1}{l|}{8.77} & \multicolumn{1}{l|}{3.5}  & 40.00 \\ 
vq-wav2vec                & 34.15M                  & \multicolumn{1}{l|}{8.47} & \multicolumn{1}{l|}{4.2}  & 73.25 \\ 
wav2vec 2.0 Base          & 95.04M                  & \multicolumn{1}{l|}{9.03} & \multicolumn{1}{l|}{3.2}  & 27.00 \\ 
HuBERT Base               & 94.68M                  & \multicolumn{1}{l|}{9.19} & \multicolumn{1}{l|}{3.4}  & 23.25 \\ \hline
PPG(TIMIT)               & -                  & \multicolumn{1}{l|}{8.32} & \multicolumn{1}{l|}{12.7}  & 84.25 \\ 
S2VC               & -                  & \multicolumn{1}{l|}{-} & \multicolumn{1}{l|}{12.4}  & 71.50 \\ \hline
Disen-TF-Codec-256bps           &    1.90M                     & \multicolumn{1}{l|}{8.60} & \multicolumn{1}{l|}{22.6}  & 64.50  \\ 
Disen-TF-Codec-256bps (vqin)           &    1.90M                     & \multicolumn{1}{l|}{8.92} & \multicolumn{1}{l|}{4.7}  & 42.00  \\ 
Disen-TF-Codec-1kbps            &     1.90M                      & \multicolumn{1}{l|}{9.08} & \multicolumn{1}{l|}{11.1}  & 42.25  \\ 
Disen-TF-Codec-256bps (IN)           &   1.90M                      & \multicolumn{1}{l|}{8.57} & \multicolumn{1}{l|}{27.2}  &  65.75 \\ 
Disen-TF-Codec-1kbps (IN)           &     1.90M                    & \multicolumn{1}{l|}{8.91} & \multicolumn{1}{l|}{11.5}  & 58.75  \\ \hline
\end{tabular}
\vspace{-0.6cm}
\end{table}
\vspace{-0.2cm}
\subsection{Comparison on Voice Conversion Benchmark}
\label{ssec:exp2}

In this section we evaluate the disentanglement ability of different features on the voice conversion benchmark as shown in Table \ref{table:2}. Besides proposed schemes, other numbers are from the S3PRL-VC paper \cite{superb-vc}. We take the learned discrete content representation $\hat{X}_{c}$ as the linguistic feature to perform VC. Taking compression as our ``pretext'' task, we compare our model with modern self-supervised learning(SSL) models. As shown in Table \ref{table:2}, most SSL models fail to convert the speaker identity, resulting in a very low ASV score, indicating that existing SSL models have weak ability to disentangle speaker from content except vq-wav2vec. vq-wav2vec \cite{vq-wav2vec} performs well on VC due to the discretization in it's archiecture, which uses an information bottleneck to enforce the model to drop global information. 

With the two-branch feature learning paradigm, our Disen-TF-Codec achieves an acceptable balance of speaker similarity and speech intelligibility and a good signal reconstruction, indicating that $\hat{X}_c$ extracted by the content encoder are relatively compact with little speaker information. It is worth noting that our Disen-TF-Codec is far more light-weight than modern SSL models with only 1.9M params for content encoder and low latency by causal implementation. Besides, the dataset used for training is much smaller than other SSL models, indicating the great potential of our coding framework. It can also be seen that the ``Disen-TF-Codec-256bps (vqin)'', that takes features before quantization as the linguistic feature for VC, performs much worse than ``Disen-TF-Codec-256bps'' in speaker similarity, which is consistent with what we observe in Figure 3 (c) and (d).



\vspace{-0.3cm}
\section{Conclusions}
\label{sec:conclusion}
\vspace{-0.3cm}
In this work, we introduce disentangled feature learning into real-time neural speech coding in low bitrate scenarios. Specifically, more global-like speaker identity and local content features are learned with disentanglement to represent speech.  Both subjective and objective experiments demonstrate that disentanglement improves the coding efficiency and enables voice conversion in compressed domain of neural speech codecs. In the future, we will investigate online normalization for better disentanglement in multi-speaker case and more detailed representations in terms of not only speaker and content information but also the prosody and emotions.

\vfill\pagebreak

\bibliographystyle{IEEEbib}
\bibliography{strings,refs}

\end{document}